\documentclass[pre,amsmath,amssymb,preprint,superscriptaddress]{revtex4}

\usepackage{amsmath}    
\usepackage{amsfonts}
\usepackage{amssymb}
\usepackage{graphicx}   
\usepackage{dcolumn}
\usepackage{bm}

\begin{document}

\title{Hierarchy of instabilities for two counter-streaming magnetized pair beams: influence of field obliquity}

\author{A. Bret}
\affiliation{ETSI Industriales, Universidad de Castilla-La Mancha, 13071 Ciudad Real, Spain}
 \affiliation{Instituto de Investigaciones Energ\'{e}ticas y Aplicaciones Industriales, Campus Universitario de Ciudad Real,  13071 Ciudad Real, Spain.}

\author{M.E. Dieckmann}
\affiliation{Department of Science and Technology (ITN), Link\"{o}pings University, Campus
Norrk\"{o}ping, SE-60174 Norrk\"{o}ping, Sweden}

\date{\today }

\begin{abstract}
The hierarchy of unstable modes when two counter-streaming pair plasmas interact over a flow-aligned magnetic field has been recently investigated [PoP \textbf{23}, 062122 (2016)]. The analysis is here extended to the case of an arbitrarily tilted magnetic field. The two plasma shells are initially cold and identical. For any angle $\theta \in [0,\pi/2]$ between the field and the initial flow, the hierarchy of unstable modes is numerically determined in terms of the initial Lorentz factor of the shells $\gamma_0$, and the field strength as measured by a parameter denoted $\sigma$. For $\theta=0$, four different kinds of mode are likely to lead the linear phase. The hierarchy simplifies for larger $\theta$'s, partly because the Weibel instability can no longer be cancelled in this regime. For $\theta>0.78$ (44$^\circ$) and in the relativistic regime, the Weibel instability always govern the interaction. In the non-relativistic regime, the hierarchy becomes $\theta$-independent because the interaction turns to be field-independent. As a result, the two-stream instability becomes the dominant one, regardless of the field obliquity.
\end{abstract}

\maketitle

\section{Introduction}
Counter-streaming instabilities in pair plasmas play a key role as the trigger of collisionless shock formation in such media \cite{SilvaApJ,BretPoP2013,BretPoP2014}. Because many instabilities such as two-stream, Weibel or oblique, compete in the linear regime, it is important to know which one grows the most in terms of the system parameters. For the unmagnetized case, counter-streaming pair plasmas are equivalent to counter-streaming electron beams, for which the instability hierarchy map has been established in terms of the beams temperatures, Lorentz factor and density ratio \cite{BretPRL2008,BretPoPReview}.

Progresses are slower for the magnetized case, due to the complexity of the analytical calculations involved when implementing a relativistic kinetic theory. The search for the most unstable mode requires sweeping the full \textbf{k}-spectrum, rendering inoperative the simplifications which can be done when the wave-vector is parallel or normal to the flow. To our knowledge, this kind of relativistic calculation has only been performed twice in the literature \cite{timofeev2009,Timofeev2013}.

In a recent paper, the hierarchy map of unstable modes has been derived for two cold colliding symmetric pair plasmas over a flow-aligned magnetic field \cite{bretPoP2016}. For such a system, the linear spectrum only depends on 2 parameters which are the strength of the field and the initial Lorentz factor of the shells. The goal of this article is to extend this previous work to an arbitrary obliquity between the flow and the field.

\begin{figure}
  \begin{center}
   \includegraphics[width=.5\textwidth]{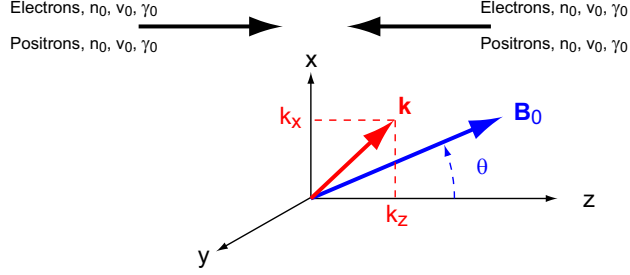}
  \end{center}
  \caption{System considered. Two  counter-streaming pair beams over an oblique magnetic field $\mathbf{B}_0$. The beams are initially cold and symmetric, with electronic density $n_0$ and positronic density $n_0$.}\label{fig:setup}
 \end{figure}

The system considered is pictured on Fig. \ref{fig:setup}. The initial flow is along the $z$ axis, and the field $\mathbf{B}_0$ lies in the $(z,x)$ plane, with $(\widehat{\mathbf{B}_0,\mathbf{e}_z})=\theta$. The wave-vector $\mathbf{k}$ of the perturbations applied to the system lies in the same $(z,x)$ plane. We therefore implement a 2D model for a direct comparison with 2D PIC simulations of such systems. The most general case would require considering an $k_y \neq 0$ component for $\mathbf{k}$. Nevertheless, previous 3D studies of the Weibel instability ($k_z=0$) found that the maximum growth-rates are to be found precisely for $k_y=0$ \cite{bretPoP2014a,Stockem2016}. The present choice of the wave-vector orientation is therefore likely to render the largest growth-rates over the full 3D \textbf{k}-space.

The same 4-fluids model than in Ref. \cite{bretPoP2016} is implemented, where 2 fluids stand for the electrons and positrons of one beam, and 2 more fluids for the electrons and positrons of the other beam. For the present article to be self-contained, the main lines of the calculation are reproduced here. We write the 4 matter conservation equations,
\begin{equation}\label{eq:matter}
  \frac{\partial n_i}{\partial t} + \nabla \cdot (n_i \mathbf{v}_i) =0,
\end{equation}
for $i=1\ldots 4$, and the 4 momentum conservation equations,
\begin{equation}\label{eq:momentum}
  \frac{\partial \mathbf{p}_i}{\partial t} + (\mathbf{v}_i\cdot\nabla) \mathbf{p}_i = q_i \left( \mathbf{E} + \frac{\mathbf{v}_i \times (\mathbf{B} + \mathbf{B}_0)}{c} \right),
\end{equation}
again with $i=1\ldots 4$. These equations are then linearized considering small departures from the initial conditions of the form $\exp(i \mathbf{k}\cdot \mathbf{r} - i \omega t)$. The first order density perturbations $n_{1i}$ are derived from Eq. (\ref{eq:matter}). Inserting them in Eq. (\ref{eq:momentum}) and using $\mathbf{B}_1 = (c/\omega)\mathbf{k}\times \mathbf{E}_1$ allows to express the first order velocity perturbations $\mathbf{v}_{1i}$ as a function of the first order field $\mathbf{E}_1$. This allows to write the first order current as,
\begin{equation}\label{eq:J1}
  \mathbf{J}_1 = \sum_{i=1}^4 q_i n_0 \mathbf{v}_{1,i} + \sum_{i=1}^4 q_i n_{1,i} \mathbf{v}_{0,i} \equiv \mathbf{J}_1 (\mathbf{E}_1).
\end{equation}
This expression is then inserted into a combination of Maxwell-Faraday's and Maxwell-Amp\`{e}re's equations, yielding
\begin{equation}\label{eq:Max}
 \mathbf{k} \times (\mathbf{k} \times \mathbf{E}_1) + \frac{\omega^2}{c^2}\left(\mathbf{E}_1 + \frac{4 \imath \pi}{\omega} \mathbf{J}_1 \right) \equiv \mathcal{T} (\mathbf{E}_1) = 0.
\end{equation}

The dielectric tensor $\mathcal{T}$ has been computed analytically with the \emph{Mathematica} tensor associated with this article and described in Ref. \cite{BretCPC}. The dispersion equation is a 16 degree polynomial which has been transferred to \emph{Matlab} for numerical analysis using a \emph{Mathematica} Notebook described in Ref. \cite{BretMM2010}. It is expressed in terms of the dimensionless parameters,
\begin{equation}\label{eq:var}
x = \frac{\omega}{\omega_p},~~\mathbf{Z} = \frac{\mathbf{k} v_0}{\omega_p},~~\beta = \frac{v_0}{c},~~\gamma_0 = \frac{1}{\sqrt{1-\beta^2}},~~\sigma = \frac{B_0^2/4\pi}{\gamma_0 (2n_0) m c^2},
\end{equation}
with
\begin{equation}
\omega_p = \sqrt{\frac{4 \pi n_0 q^2}{m}}.
\end{equation}
Note that the magnetic field is measured through $\sigma$ in Eq. (\ref{eq:var}), which is the parameter typically used in collisionless shocks physics \cite{Marco2016}. The factor ``2'' at the denominator aims at accounting for the total kinetic energy of the shells. Some authors sometimes use the parameter $\epsilon_B$ instead of $\sigma$, with $\epsilon_B=\sigma/2$ \cite{SilvaApJ,Sironi2011ApJ}.

An important point for the present study is that the state which is perturbed is not an equilibrium for $\theta\neq 0$. As  emphasized in previous studies \cite{BretPoPOblique,BretPoP2013a,bretPoP2014a}, the forthcoming calculations are valid as long as the response of the system to growing harmonic perturbations, is faster than its response to the tilted magnetic field. Since the latter response develops on time scales comparable to the cyclotron frequency of the charges, while the former response follows the growth-rate $\Delta$, we need,
\begin{equation}
\Delta > \frac{q B_0 \sin\theta}{\gamma_0 m c},
\end{equation}
where the $\sin\theta$ factor accounts for the fact that only the normal component of the field is involved in this respect. Dividing both sides by $\omega_p$ gives,
\begin{equation}\label{eq:validity}
\frac{\Delta}{\omega_p} \equiv \delta > \sin\theta \sqrt{\frac{2\sigma}{\gamma_0}}.
\end{equation}
The growth-rates we are about to compute are therefore relevant only if they satisfy the condition above. The threshold so defined will be numerically  computed in the sequel, and an analytical counterpart will be given when possible.

For a given set of parameters $(\gamma_0,\sigma,\theta)$, we compute the growth-rate $\delta(Z_z,Z_x)$. We then determine the ``hierarchy map'' under the form,
\begin{eqnarray}
\delta_{max}(\gamma_0,\sigma,\theta) &=& \max \{ \delta(Z_z,Z_x), (Z_z,Z_x) \in \mathbb{R}^2 \},\nonumber\\
\mathbf{Z}_{max}(\gamma_0,\sigma,\theta)      &=& (Z_z,Z_x) / \delta(Z_z,Z_x) = \delta_{max}.
\end{eqnarray}
A hierarchy map gives therefore the most unstable $\mathbf{Z}$ and its growth-rate for any set of parameters $(\gamma_0,\sigma,\theta)$.

  \begin{figure}
  \begin{center}
   \includegraphics[width=\textwidth]{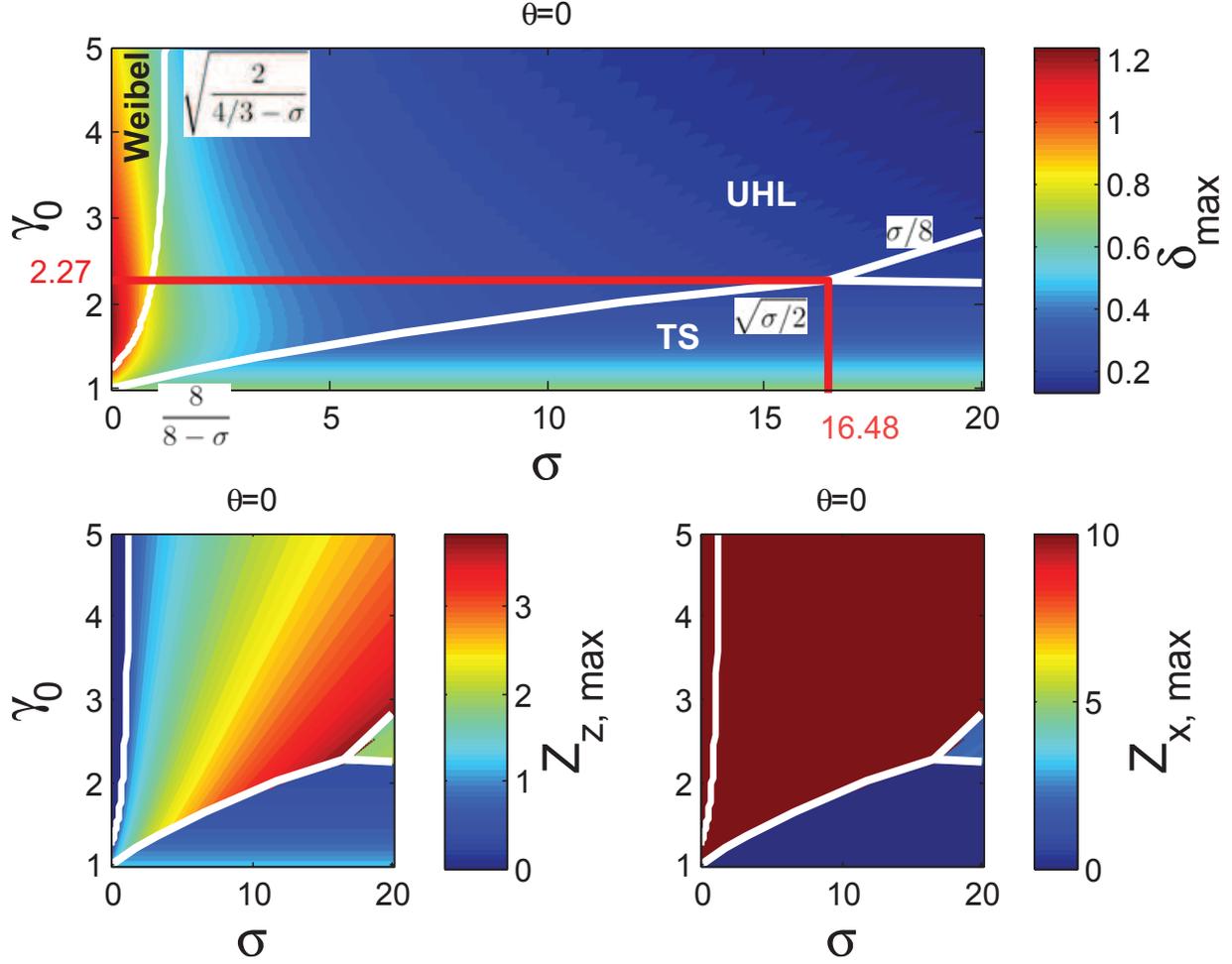}
  \end{center}
  \caption{Hierarchy map for $\theta=0$. This figure is the counterpart of Figs. 1 \& 7 of Ref \cite{bretPoP2016}, with the mapping (\ref{eq:map}). The upper-plot shows the largest growth-rate  for any given couple $(\sigma,\gamma_0)$. The lower-left plot shows the $Z_z$ component of the most unstable $\mathbf{Z}$, and the lower-right plot shows its $Z_x$ component. The 4 modes governing the system are Weibel, two-stream (TS), oblique and upper-hybrid-like \cite{Godfrey1975} (UHL). }\label{fig:theta0}
 \end{figure}

\section{Hierarchy map for $\theta=0$}
The case $\theta=0$ has been explored in Ref. \cite{bretPoP2016}. However, the field strength was parameterized by the $\Omega_B$ parameter, different from $\sigma$ given in Eqs. (\ref{eq:var}). The correspondence between them reads,
\begin{equation}\label{eq:map}
\Omega_B = \frac{1}{\omega_p}\frac{q B_0}{m c} ~~ \Rightarrow   ~~  \Omega_B = \sqrt{2\gamma_0\sigma}.
\end{equation}
We therefore display on Fig. \ref{fig:theta0} the hierarchy map for $\theta=0$ in terms of $\gamma_0$ and $\sigma$. The equations for the frontiers are here translated from Fig. 7 of Ref. \cite{bretPoP2016} to the new mapping.

In the cold limit, some modes, like the Weibel modes for example, have their growth-rate saturating at large $Z_x$ \cite{califano3,BretPoPReview}. When such a mode governs the spectrum, the corresponding $Z_{x,max}$ is set to 10. We recover a hierarchy map governed by 4 different kind of modes, with a triple point at $(\sigma,\gamma_0)=(16.48,2.27)$.

We now turn to the hierarchy map for $\theta=\pi/2$ before exploring intermediate obliquities.

  \begin{figure}
  \begin{center}
   \includegraphics[width=\textwidth]{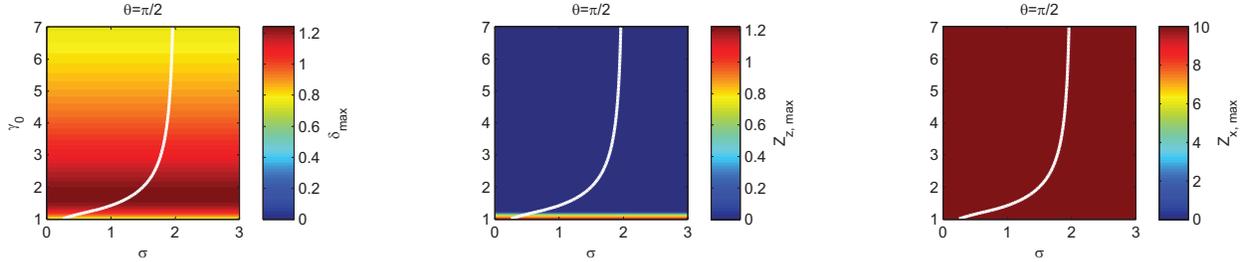}
  \end{center}
  \caption{Hierarchy map for $\theta=\pi/2$. The white line has been numerically determined, and shows where equality is reached for the validity condition (\ref{eq:validity}). Calculations are valid to the left of this line which intersects the axis $\gamma_0=1$ for $\sigma=1/4$ (see Eq. \ref{eq:validpisur2}).}\label{fig:thetapisur2}
 \end{figure}

\section{Hierarchy map for $\theta=\pi/2$}
The result of the numerical computation of the hierarchy map for $\theta=\pi/2$ is displayed on Fig. \ref{fig:thetapisur2}. The validity condition (\ref{eq:validity}) arising from the field obliquity is fulfilled to the left of the white line, which intersects the axis $\gamma_0=1$ for $\sigma=1/4$ (see Eq. \ref{eq:validpisur2} below). For the sake of the figure, it has been numerically determined although we derive below its exact analytical expression.

The computation unravels an extremely simple hierarchy map: it simply does not depend on $\sigma$. Previous works already found that the Weibel growth-rate becomes independent of $\sigma$ for $\theta=\pi/2$ because the instability has the charges moving sideways, that is, parallel to the field \cite{Stockem2016}. Other $\sigma$-dependent modes grow (not shown), but they do not outgrow the Weibel mode.

The hierarchy map shows that while the fastest growing mode always has $Z_x=10$, that is $Z_x=\infty$, the $Z_z$ component is not always 0, as expected for Weibel. We now discuss this point analytically by solving the problem in the limit $Z_x=\infty$. The dispersion function $P$ is here a 4th degree polynomial in $Z_x$. The dispersion function $P_{Z_x\infty}$ for $Z_x=\infty$ is therefore the coefficient of $Z_x^4$ in $P$. It reads,
\begin{eqnarray}\label{eq:disperpi/2}
  P_{Z_x\infty} &=& QR^2, \\
  Q             &=& -16 \beta^2 - 4(x^2+Z_z^2)/\gamma_0+ (x^2-Z_z^2)^2\gamma_0^2 \nonumber \\
  R             &=& 4 \sigma ^2+\gamma_0^6 \left(x^2-Z_z^2\right)^2-4\gamma_0^3 \sigma  (x^2+Z_z^2) \nonumber \\
\end{eqnarray}
Some simple algebra shows that the equation $R=0$ does not yield any unstable mode. We are thus left with $Q=0$, which, as expected from Fig. \ref{fig:thetapisur2}, is independent of $\sigma$. Solving $Q=0$ gives,
\begin{equation}\label{eq:tauxpisur2}
\delta^2 = \frac{2+\gamma_0^3 Z_z^2-2 \sqrt{4 \beta ^2 \gamma_0^4+2 \gamma_0^3 Z_z^2+1}}{\gamma_0^3},
\end{equation}
with $\delta^2 < 0$ for $Z_z < 2/\sqrt{\gamma_0}$. The growth-rate $\delta$ reaches its maximum for,
\begin{eqnarray}\label{eq:Zzmaxpisur2}
Z_{z,max} &=& \frac{\sqrt{3/2-2 \gamma_0^4+2 \gamma_0^2}}{\gamma_0^{3/2}},~~\mathrm{for}~~\gamma_0 < \sqrt{3/2}, \nonumber\\
Z_{z,max} &=& 0, ~~\mathrm{for}~~ \gamma_0 > \sqrt{3/2}
\end{eqnarray}
because the expression under the square-root turns negative for $\gamma_0 > \sqrt{3/2}$. The maximum growth-rate then reads,
\begin{eqnarray}\label{eq:Tauxmaxpisur2}
\delta_{max} &=& (1+\beta^2)\sqrt{\frac{\gamma_0}{2}},~~\mathrm{for}~~ \gamma_0 < \sqrt{3/2}, \nonumber\\
\delta_{max} &=& 2\frac{\beta}{\sqrt{\gamma_0}}, ~~\mathrm{for}~~ \gamma_0 > \sqrt{3/2}
\end{eqnarray}

  \begin{figure}
  \begin{center}
   \includegraphics[width=\textwidth]{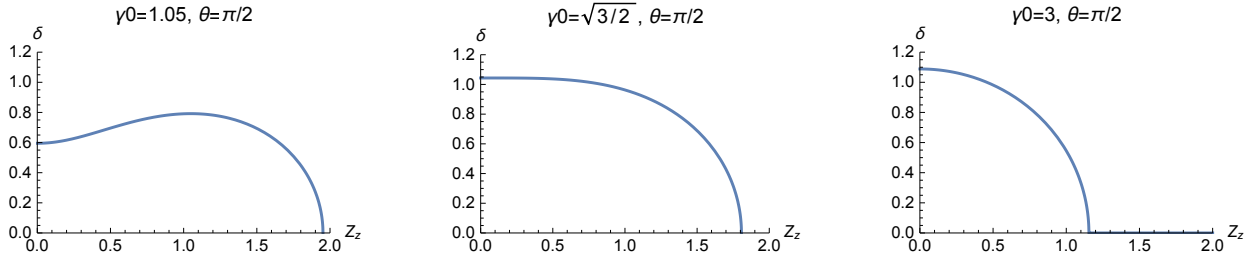}
  \end{center}
  \caption{Growth-rate (\ref{eq:tauxpisur2}). For $\theta=\pi/2$, the maximum growth-rate is given by a dispersion equation which does not depend on $\sigma$. At any rate, the dominant mode has $Z_{x,max}=\infty$. For $\gamma_0 > \sqrt{3/2}$, $Z_{z,max}=0$ and the Weibel instability governs the linear phase.}\label{fig:Wthetapisur2}
 \end{figure}

These different regimes are illustrated on Fig. \ref{fig:Wthetapisur2}, explaining the hierarchy map (\ref{fig:thetapisur2}). The validity condition (\ref{eq:validity}) reads here,
\begin{eqnarray}\label{eq:validpisur2}
  \gamma_0 &>& \sqrt{\frac{2}{2-\sigma }}, ~~\gamma_0 > \sqrt{3/2}  ~~ (\sigma > 2/3) \nonumber\\
  \gamma_0 &>& \frac{\sqrt{\sigma +\sqrt{\sigma  (\sigma +2)}+1}}{\sqrt{2}}, ~~\gamma_0 < \sqrt{3/2}~~ (\sigma < 2/3),
\end{eqnarray}
where the validity threshold $\sigma<1/4$ is analytically recovered for $\gamma_0=1$, together with its vertical asymptote at $\sigma=2$.

  \begin{figure}
  \begin{center}
   \includegraphics[width=\textwidth]{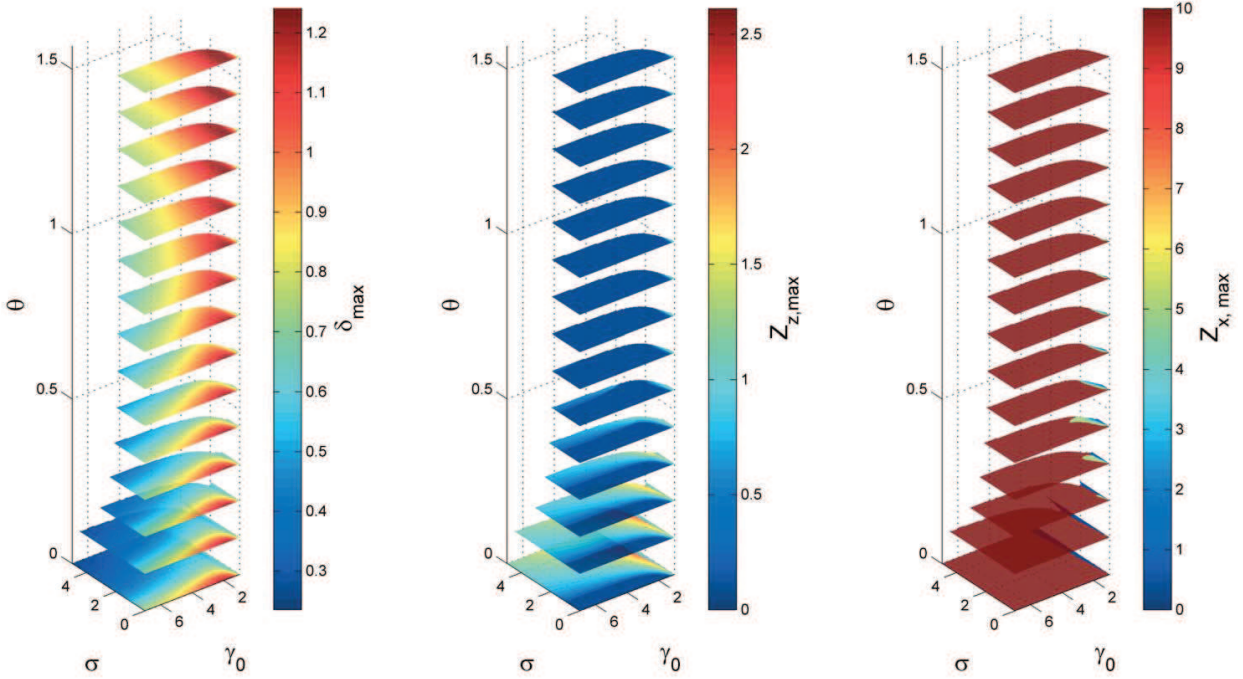}
     \end{center}
  \caption{Hierarchy maps for intermediate obliquities. Only the part of the maps satisfying the validly condition (\ref{eq:validity}) is displayed. It has been numerically determined. An alternative presentation of these maps is available in the Supplementary Material file ``Hierarchies.pdf''.}\label{fig:inter}
 \end{figure}

\section{Hierarchy maps for intermediate obliquities}
While the hierarchy map for $\theta=0$ is quite involved, the one for $\theta=\pi/2$ is remarkably simple. We now examine how one map evolves into the other when the field obliquity  progressively increases from $\theta=0$ to $\pi/2$.

Figure \ref{fig:inter} pictures the numerical calculation of the hierarchy maps for intermediate values of $\theta$, together with the domain where condition (\ref{eq:validity}) is fulfilled for each (also numerically determined). An alternative presentation of these maps is available in the Supplementary Material file ``Hierarchies.pdf''. A few points are worth commenting:

\begin{itemize}
  \item In the relativistic regime, the fastest mode always has $Z_x=\infty$. Yet, its $Z_z$ component may be 0 (Weibel), or finite. We will therefore study the border between these two regimes.
  \item Around $\theta=0.45$ $(25^\circ)$, we find a region of the phase space centered around $(\sigma,\gamma_0) \sim (1,1.7)$ with both $Z_x \neq 0$ and $Z_z \neq 0$. We will comment on this oblique modes regime.
  \item The non-relativistic regime can be analyzed analytically because it is  equivalent to the field-free limit. We will therefore comment it in details.
\end{itemize}

\subsection{Relativistic regime}
The hierarchy maps indicate that in the relativistic regime, the fastest mode always has $Z_x=\infty$. We thus derive the dispersion equation in this limit. Its exact expression is reported in Appendix \ref{ap:GdZx}.

Expanding it in powers of $\gamma_0$ and keeping the three largest powers, namely 10, 9 and 8, gives the following approximate dispersion equation in the $Z_x=\infty$ and large $\gamma_0$ limit,
\begin{equation}\label{eq:DisperGdZxGdGam}
  \gamma_0^8 (x^2-Z_z^2)^2 \left(4 \sigma ^2 \cos ^4\theta+\gamma_0^2 (x^2-Z_z^2)^2-4 \gamma_0 \sigma  \cos^2\theta (x^2+Z_z^2)-16\right) = 0,
\end{equation}
which solution yielding unstable modes gives,
\begin{equation}
\delta^2 \sim \frac{\sigma  \cos (2 \theta )+\sigma +\gamma_0 Z_z^2 -2 \sqrt{2 \gamma_0 \sigma  \cos^2\theta Z_z^2+4}}{\gamma_0},
\end{equation}
which is not an exact expression due to our large $\gamma_0$ expansion.

The profile $\delta(Z_z)$ is still similar to the one pictured on Fig. \ref{fig:Wthetapisur2}. Hence, the largest $\delta$ will have $Z_z =0$ or not, depending on the sign of $\partial^2\delta/\partial Z_z^2$ in $Z_z=0$. A little algebra shows this second derivative vanishes in $Z_z=0$ for,
\begin{equation} \label{eq:WregimeOblique}
\sigma \sim 1/\cos^2\theta.
\end{equation}
We thus find that as expected from Fig. \ref{fig:inter}, the level of magnetization required to exit the Weibel regime increases with the field obliquity.

Another feature of Fig.  \ref{fig:inter} is that for large $\gamma_0$'s, condition (\ref{eq:validity}) translates to a condition of the type $\sigma<\sigma_T(\theta)$, where $\sigma_T$ is a magnetization threshold beyond which the present calculations are no longer relevant.

We can therefore expect that beyond a certain obliquity $\theta_W$, the critical sigma given by Eq. (\ref{eq:WregimeOblique}) becomes larger than $\sigma_T$, so that the Weibel instability governs the relativistic regime whenever the present theory is relevant.

An analytical determination of $\theta_W$ has not been possible. Yet, examining Fig. \ref{fig:inter} shows $\theta_W \sim 0.78$ (44$^\circ$). While for $\theta \lesssim 44^\circ$, various modes are likely to govern the linear phase, the Weibel instability always govern the relativistic unstable spectrum for $\theta \gtrsim 44^\circ$.

%
%
%

  \begin{figure}
  \begin{center}
   \includegraphics[width=\textwidth]{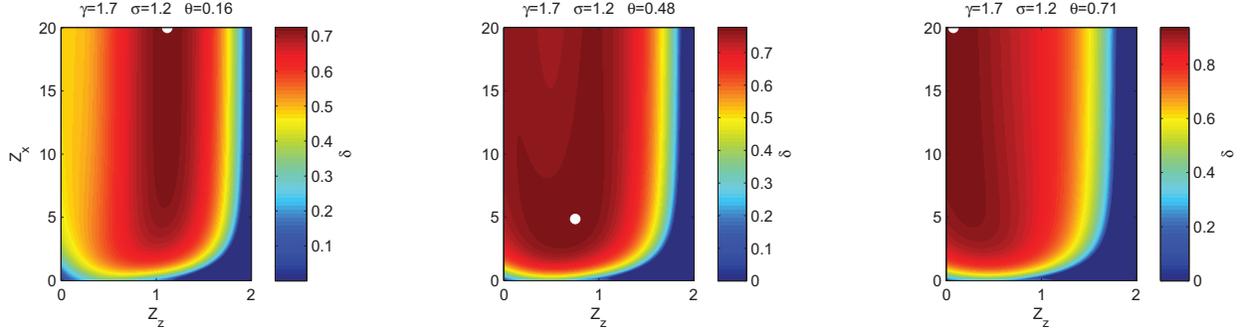}
  \end{center}
  \caption{Calculations of $\delta (Z_z,Z_x)$ for $\sigma=1.2$, $\gamma_0=1.7$ and 3 values of $\theta$ bracketing $\theta=0.48$ (25$^\circ$). The white spot localizes the most unstable $\mathbf{Z}$.}\label{fig:obli}
 \end{figure}

\subsection{Oblique regime near $\sigma=1.2$, $\gamma_0=1.7$ and $\theta=0.48$ (25$^\circ$)}
Figure \ref{fig:inter} shows that there is a regime fully governed by oblique modes with $Z_z\neq 0$ and $Z_x\neq 0$, around $\theta=0.48$.

Figure $\ref{fig:obli}$ shows 3 calculations of $\delta (\mathbf{Z})$ for $\sigma=1.2$, $\gamma_0=1.7$ and 3 values of $\theta$ bracketing $\theta=0.48$. In each case, the white spot localizes the most unstable $\mathbf{Z}$. Although it is definitely oblique for $\theta=0.48$, the growth-rate gradient toward large $Z_x$'s is extremely weak. The $Z_x$ component of the fastest growing mode switches therefore quickly from a finite value to $Z_x=\infty$. Such is the reason why the transition away from this oblique regime looks discontinuous on the hierarchy maps. It is indeed continuous, but extremely steep.

\subsection{Non-relativistic regime}
A series of hierarchy maps for various intermediate orientations has been computed emphasizing the non-relativistic regime $\beta \ll 1$. The results are displayed in the Supplementary Material file ``Hierarchies\_NR.pdf''.

A clear picture emerges: in this regime, the $Z_z$ component of the fastest growing mode $\mathbf{Z}_{max}$, together with its growth-rate, settles to some constant values with $Z_{z,max} \sim 1.2$ and $\delta_{max} \sim 0.7$. As for the $Z_x$ component of $\mathbf{Z}_{max}$ its behaviour is not so monotonic.

These patterns can be understood noting that the Lorenz force $\mathbf{F}_L$, through which the field $\mathbf{B}_0$ acts on the system, decreases with $\beta$ because $F_L \propto \beta B_0$. Since the maximum $B_0$ we can explore is bounded by the validity condition (\ref{eq:validity}), the non-relativistic limit is also the field-free, $B_0=0$, limit.

What is then the non-relativistic behavior of the system for $B_0=0$? First, it becomes independent of $\theta$ since there is no longer any $\mathbf{B}_0$, and second, the dominant instability is the two-stream instability \cite{BretPRL2008,BretPoPReview}.

Setting therefore $B_0=0$, we derive the dispersion equation for the two-stream instability,
\begin{equation}\label{eq:ts}
  1-\frac{2}{(x+Z_z)^2}-\frac{2}{(x-Z_z)^2} = 0.
\end{equation}
Solving for $x$ gives the exact growth-rate,
\begin{equation}\label{eq:TauxTS}
\delta_{TS}^2 = 2 + Z_z^2-2 \sqrt{2 Z_z^2+1},
\end{equation}
reaching its maximum,
\begin{equation}\label{eq:TauxmaxTS}
\delta_{TS,max} = \frac{1}{\sqrt{2}} \sim 0.7,
\end{equation}
for,
\begin{equation}\label{eq:ZzTS}
Z_{z,max} = \sqrt{\frac{3}{2}} \sim 1.22.
\end{equation}
We thus find that these $Z_{z,max}$ and $\delta_{TS,max}$ fit precisely what is observed in the hierarchy maps.

  \begin{figure}
  \begin{center}
   \includegraphics[width=\textwidth]{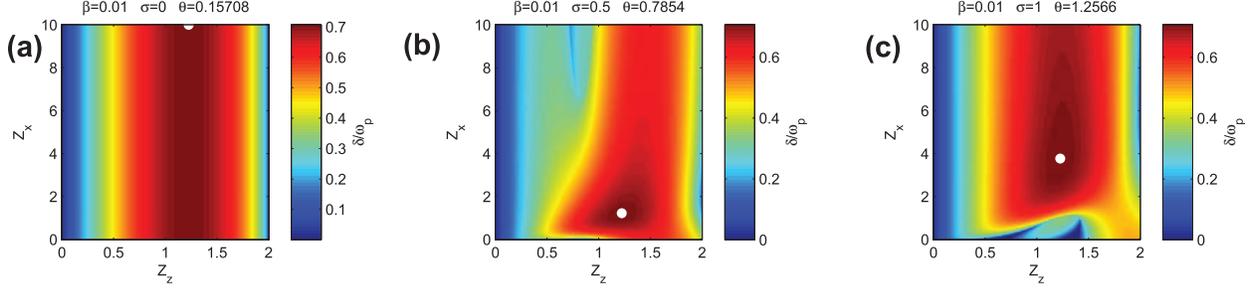}
  \end{center}
  \caption{Three maps $\delta(\mathbf{Z})$ for $\beta=10^{-2}$, and various combinations of $(\sigma,\theta)$, the first one having $\sigma=0$. All maps tend to map (a) in the $\beta \rightarrow 0$ limit. The white point shows the location of the fastest growing mode.}\label{fig:NR}
 \end{figure}

Let us finally turn to the behavior of $Z_{x,max}$. Figure \ref{fig:NR} displays three maps $\delta(\mathbf{Z})$ for $\beta=10^{-2}$, and various combinations of $(\sigma,\theta)$, the first one having $\sigma=0$. This latter map has been known for long \cite{Watson,Bludman,fainberg}. The two-stream instability is found for $Z_x=0$, and there is nearly no $Z_x$ dependance of the growth-rate.

Since for $\sigma=0$ and $\beta \ll 1$, neither the maximum growth-rate, nor its location along the $Z_z$ axis, depend on $\beta$ [see Eqs. (\ref{eq:TauxmaxTS},\ref{eq:ZzTS})], this map is the limit of the growth-rate when $\beta \rightarrow 0$. As a consequence, any map $\delta(\mathbf{Z},\sigma,\theta,\beta)$ has to tend to this one when $\beta \rightarrow 0$. The convergence is simple for $\delta_{max}$ and $Z_{z,max}$. However, the behavior of $Z_{x,max}$ is more involved. The reason for this is that $Z_{x,max}$ must definitely tend to $\infty$, and indicated by Fig. \ref{fig:NR}(a), but the gradient $\partial\delta_{max}/\partial Z_x$ in nearly zero. Therefore, while we still have $\lim_{\beta \rightarrow 0}Z_{x,max} = \infty$, the path to the limit is much more varied that it is for the $Z_{z,max}$ component.

Although temperature effects are beyond the scope of this work, they are likely to ``regularize'' the convergence of $Z_{x,max}$. The reason for this is that temperature always tend to stabilize large $\mathbf{k}$'s because pressure opposes the formation of small density bumps \cite{Silva2002,BretPoPMagne}. For the field-free case, it has been found that this effect localizes $Z_{x,max}$ to a finite value \cite{BretPRL2005}. The same has been found for the magnetized case  \cite{timofeev2009,Timofeev2013}. Hence, for realistic settings, or even in the case of small temperatures, the convergence of $Z_{x,max}$ is likely to resemble the one of $Z_{z,max}$.

\section{Conclusion}
We have considered a counter-streaming pair plasmas system, over an arbitrarily tilted magnetic field. The two colliding plasma shells are initially both cold and of the same density. The unstable linear spectrum is determined by only 3 parameters: the initial Lorentz factor of the shells $\gamma_0$, the strength of the magnetic field as measured by the $\sigma$ parameter, and its obliquity $\theta$.

The wave-vector of the perturbations pertains to the plane where the fastest growing modes are expected. For field obliquities varying from $0$ to $\pi/2$, we numerically determined the hierarchy map of linear instabilities, namely, the fastest growing mode $\mathbf{Z}_{max}(\sigma,\gamma_0)$ and its growth-rate $\delta_{max}(\sigma,\gamma_0)$.

The case $\theta=0$ has already been treated in Ref. \cite{bretPoP2016} and shows 4 different kinds of modes are likely to govern the linear phase. For finite obliquities, the hierarchy maps tends to simplify for 2 reasons:
\begin{enumerate}
  \item For these calculations to be relevant, the growth-rate must be larger than the cyclotron frequency of the charges in the field component normal to the initial flow. This sets condition (\ref{eq:validity}), which restricts the relevant $(\sigma,\gamma_0)$ domain.
  \item While the Weibel instability can be perfectly cancelled for $\theta=0$ \cite{Stockem2006ApJ}, it cannot for $\theta \neq 0$ \cite{BretPoP2011,BretPoP2013a,bretPoP2014a}. When $\theta=\pi/2$ is reached, it is simply unaffected because it has the particles moving sideways, that is, parallel to the field. As a result, this instability gains robustness and tend to govern a larger part of the map.
\end{enumerate}

In the relativistic regime, upper-hybrid-like (UHL) and Weibel modes mostly share the map until $\theta \sim 0.78$ (44$^\circ$). Beyond this critical obliquity, the UHL domain is ruled out by the validity condition (\ref{eq:validity}), and the Weibel instability always governs.

In the non-relativistic limit, the map progressively becomes $\theta$-independent simply because the dynamics becomes field-independent. In such conditions, the interaction is governed by the two-stream instability, even if the normal component of the most unstable $\mathbf{Z}$ may tend to 0 is a complicated way.

Our results show that the Weibel instability still dominates for $\sigma = 0.1$, while  Ref. \cite{Sironi2009Apj} suggests it does not. This seeming contradiction can be resolved by a comparison of the assumptions taken by us and in Ref. \cite{Sironi2009Apj}. We assume that we are in the reference frame in which the magnetic field is at rest and where the total momentum of the cold beams vanish. We analyze the instability spectrum in the full $\mathbf{k}$-space. The initial conditions used in Ref. \cite{Sironi2009Apj} introduce an oblique magnetic field into a moving pair cloud. The magnetic field has a component along the plasma's propagation direction and one that points out of the two-dimensional simulation plane. A filamentation of the pair cloud along the direction of the magnetic field is excluded by the simulation geometry. However, this is the direction along which the instability would develop, because the magnetic field does not affect the particle mobility along its direction. Reference \cite{Sironi2009Apj} examines the shock formation also in 3 spatial dimensions, but only for the quasi-parallel magnetic field directions that let the filamentation instability grow also in the 2D simulations. The rapid heating of the inflowing upstream plasma furthermore implies that the interacting beams cease to be cold in the simulation after a short time. We do not take into account thermal effects in our analysis.

However, the main reason for the difference between the results obtained by us and those of Ref. \cite{Sironi2009Apj} could be that in the simulation, the magnetic field  is moving with the pair cloud. This is achieved by the introduction of a convective electric field. The pair cloud that is reflected by the wall has the same density, speed modulus and temperature than the inflowing plasma. This symmetry implies that the magnetic field in the beam overlap layer, which is occupied by the inflowing and by the reflected particles, has to be at rest in the reference frame of the wall; the convective electric field in this interval must vanish. The convective electric field points along the simulation direction, which is orthogonal to the beam direction, and its amplitude varies along the beam direction. This electric field is thus rotational, which lets a magnetic field grow at the interface between the inflowing pair cloud and the overlap layer. The energy density of the initial magnetic field is not small compared to the particle's kinetic energy and its amplification will extract a substantial amount of energy from the upstream plasma. We neglect this important nonlinear effect in our analytic work.

In addition, the instability which is first triggered when the two shells overlap may not be the one eventually forming, and/or sustaining the shock. Such in the case for example when two electron/ion plasmas collide \cite{Bret2015ApJL}. In that case, electrons turn Weibel unstable first, but the field they grow is found unable to form a shock. It takes the ions to turn Weibel unstable, and then their Weibel filaments to merge, in order to start the shock formation. For the present case, further works will be needed to discriminate the instability first triggered, from the mechanism responsible for the shock formation and/or sustainment.

\section{Acknowledgments}
This work was supported by grants ENE2013-45661-C2-1-P and ENE2016-75703-R from the Ministerio de Educaci\'{o}n y Ciencia, Spain and grant PEII-2014-008-P from the Junta de Comunidades de Castilla-La Mancha. The authors thank Lorenzo Sironi and Asaf Pe'er for fruitful discussions.

\appendix
\section{Dispersion equation for $Z_x=\infty$}\label{ap:GdZx}
The exact dispersion equation for $Z_x=\infty$ reads,
\begin{eqnarray}\label{eq:DisperGdZx}
0&=&\gamma_0^6 (x^2-Z_z^2)^2 \left(16-16 \gamma_0^2+\gamma_0^4 (x^2-Z_z^2)^2-4 \gamma_0 (x^2+Z_z^2)\right) \nonumber\\
&+& 2\sigma (A + B + C - D),
\end{eqnarray}
with,
\begin{eqnarray}
A&=& 2\gamma_0^8 \sigma  \cos^4\theta (x^2-Z_z^2)^2-\gamma_0^3 \sigma \sin ^2(2\theta) \left(2(x^2+Z_z^2)-\gamma_0^3 (x^2-Z_z^2)^2\right)     \nonumber\\
B&=& 2\gamma_0^6 \cos ^2\theta (x^2-Z_z^2)^2 \left(2-\gamma_0^3 (x^2+Z_z^2)\right)                            \nonumber\\
C&=& 2\sigma  \sin^4\theta \left(16-16 \gamma_0^2+\gamma_0^4 (x^2-Z_z^2)^2-4 \gamma_0 (x^2+Z_z^2)\right) \nonumber\\
D&=& 2\gamma_0^3 \sin^2\theta (x^2+Z_z^2) \left(16-16 \gamma_0^2+\gamma_0^4 (x^2-Z_z^2)^2-4 \gamma_0 (x^2+Z_z^2)\right) \nonumber.
\end{eqnarray}


\end{document}